\documentclass[eat,twocolumn]{jmlr}
\usepackage{comment}
\usepackage{bbm}
 \usepackage{rotating}% for sideways figures and tables
\usepackage{longtable}% for long tables

\usepackage{booktabs}
\usepackage[load-configurations=version-1]{siunitx} % newer version
\theorembodyfont{\upshape}
\theoremheaderfont{\scshape}
\theorempostheader{:}
\theoremsep{\newline}

\jmlrvolume{}
\firstpageno{1}
\jmlryear{2022}
\jmlrworkshop{Machine Learning for Health (ML4H) 2022}

\title[TFT for Long-term Explainable Prediction of ED Overcrowding]{A Temporal Fusion Transformer for Long-term Explainable Prediction  of Emergency Department Overcrowding  
\titletag{\thanks{This work was partially supported by the strategic project NOVA LINCS (UIDB/04516/2020), the FCT project DSAIPA/AI/0087/2018 and the Carnegie Mellon University - Portugal FCT project CMU/TIC/0016/2021}}}

   \author{\Name{Francisco M. Caldas\nametag{}} \Email{f.caldas@campus.fct.unl.pt}\and
   \Name{Cláudia Soares} \Email{claudia.soares@fct.unl.pt}\\
   \addr NOVA School of Science and Technology, Caparica, Portugal}

\begin{document}

\maketitle

\begin{abstract}
Emergency Departments (EDs) are a fundamental element of the Portuguese National Health Service, serving as an entry point for users with diverse and very serious medical problems. Due to the inherent characteristics of the ED, forecasting the number of patients using the services is particularly challenging. And a mismatch between affluence and the number of medical professionals can lead to a decrease in the quality of the services provided and create problems that have repercussions for the entire hospital, with the requisition of healthcare workers from other departments and the postponement of surgeries.
%Emergency Departments (ED) are a key part of the complete Health Care System, playing a very import role in patient care, providing emergency health care services to people in critical situations. To better manage the inflow o 
%Unlike other hospital departments, the ED is difficult to manage, since a mismatch between personnel and patient volume poses a challenge to hospital managers, and puts at risk the quality of medical services for the entire hospital. 
ED overcrowding is driven, in part, by non-urgent patients that resort to emergency services despite not having a medical emergency, representing almost half of the total number of daily patients. This paper describes a novel deep learning architecture, the Temporal Fusion Transformer, that uses calendar and time-series covariates to forecast prediction intervals and point predictions for a 4-week period. We have concluded that patient volume can be forecast with a Mean Absolute Percentage Error (MAPE) of 9.87\%  for Portugal’s Health Regional Areas (HRA) and a Root Mean Squared Error (RMSE) of 178 people/day. The paper shows empirical evidence supporting the use of a multivariate approach with static and time-series covariates while surpassing other models commonly found in the literature.
\end{abstract}
\begin{keywords}
Time Series  
Emergency Department 
Machine Learning 
Temporal Fusion Transformer 
Forecasting 
Manchester Triage System 
Neural Network 
Explainable ML 
National Health Service
\end{keywords}

\section{Introduction}
\label{sec:intro}

%This is a sample article that uses the \textsf{jmlr} class with
%the \texttt{pmlr} class option.  Please follow the guidelines in
%this sample document as it can help to reduce complications when
%combining the articles into a book. Please avoid using obsolete
%commands, such as \verb|\rm|, and obsolete packages, such as
%\textsf{epsfig}.\footnote{See
%\url{http://www.ctan.org/pkg/l2tabu}}

%Please also ensure that your document will compile with PDF\LaTeX.
%If you have an error message that's puzzling you, first check for it
%at the UK TUG FAQ
%\url{https://texfaq.org/FAQ-man-latex}.  If
%that doesn't help, create a minimal working example (see
%\url{https://www.dickimaw-books.com/latex/minexample}) and post
%to somewhere like TeX on StackExchange
%(\url{https://tex.stackexchange.com/}) or the LaTeX Community Forum
%(\url{https://latex.org/forum/}).

%\begin{note}
%This is an numbered theorem-like environment that was defined in
%this document's preamble.
%\end{note}

The forecast of the number of patients who use emergency services daily is essential to determine in advance the human resources needed at hospital Emergency Departments (ED). Multi-step ahead predictions allow hospital managers to organise rotation schedules and diminish waiting times in urgent care facilities \citep{Wargon2009,Jones2008}. When not accounted for, overcrowding can lead to a decrease in the quality of patient care and worse clinical outcomes \citep{Bernstein2009,Hurwitz2014}. From a macro point of view, the influx in the emergency department combines an expected number of people who are taken to the emergency room with a very serious illness, for example, heart attack, with people that use the emergency hospital to deal with non urgent problems, such as common cold, strained muscles, or to deal with problems associated with chronic illness \citep{Zachariassee2019,Hurwitz2014}. The most serious cases are reasonably constant over time, 
and, predominantly, people in life threatening conditions have no choice but to go to emergency care, thus the indicators of a rise in patients with serious illnesses might not be the same for non urgent users.
A large number of patients that resort to urgent care are not, however, urgent, according to the Manchester Triage system, used in the Portuguese National Healthcare System. Roughly 40\% of the patients are classified during triage at the green/blue level, which means not urgent. Unlike more urgent patients, the influx of green/blue patients has several factors that follow well-defined cycles. For example, it is easy to identify that the day with the most influx of non-urgent patients is Monday, with a smaller number of patients pursuing emergency care during the weekend \citep{batal2001,holleman1996,rathlev2007time}. To combine the predictive power of Deep Neural Networks with the explainability usually reserved for simpler algorithms, we will use a recently developed machine learning model to predict the influx of non-urgent patients: the Temporal Fusion Transformer (TFT)~\citep{Lim2021}; and study which variables, time-series or not, had the most impact on the model, and thus which are most relevant to predict daily patient volume. 

Previous studies have examined the multi-step forecasting of daily patient volumes \citep{Jones2008,diehl1981}. Most focus is on the use of classical statistical tools for temporal linear regression such as moving averages \citep{milner1988}, and their many extensions, namely ARIMA, SARIMA or VARIMA \citep{Afilal2016,Schweigler2009,whitt2019,carvalho2018}. In recent years, with the advent of machine learning, newer studies have been conducted that use neural networks \citep{Jones2008,Zhou2018}, or otherwise other machine learning techniques to tackle the same problem \citep{Rocha2021,Navares2018,Tuominen}. From the use of Feed-forward Neural Networks \citep{Jones2008,Navares2018}, to Recurrent Neural Networks \citep{Kadri2020,Harrou2020}, 1-D Convolution Neural Networks \citep{Sharafat2021}, and later to Long Short-Term Unit (LSTM) \citep{Sudarshan2021,Harrou2020}, there has been a constant advance in the field, from linear models to deep neural network models.

In most studies using ARIMA and its variants, it was found that calendar variables (day, day of the week, holidays) have a significant contribution to model results \citep{boyle2012,hertzum2017,Jones2008,Wargon2009}. A more detailed Literature Review can be found in Annex A.

\section{Data Analysis}

\begin{figure}[htbp]
 % Caption and label go in the first argument and the figure contents
 % go in the second argument
\floatconts
  {fig:blue_white}
  {\caption{Time series from January, 2019 to February 20, 2020. We can observe the weekly cycle, as well as annual trends and volume variation according to RHA. %The weekends are marked with grey lines, corresponding with diminishing number of non urgent patients searching for emergency care.
  }}
  {\includegraphics[width=0.9\linewidth]{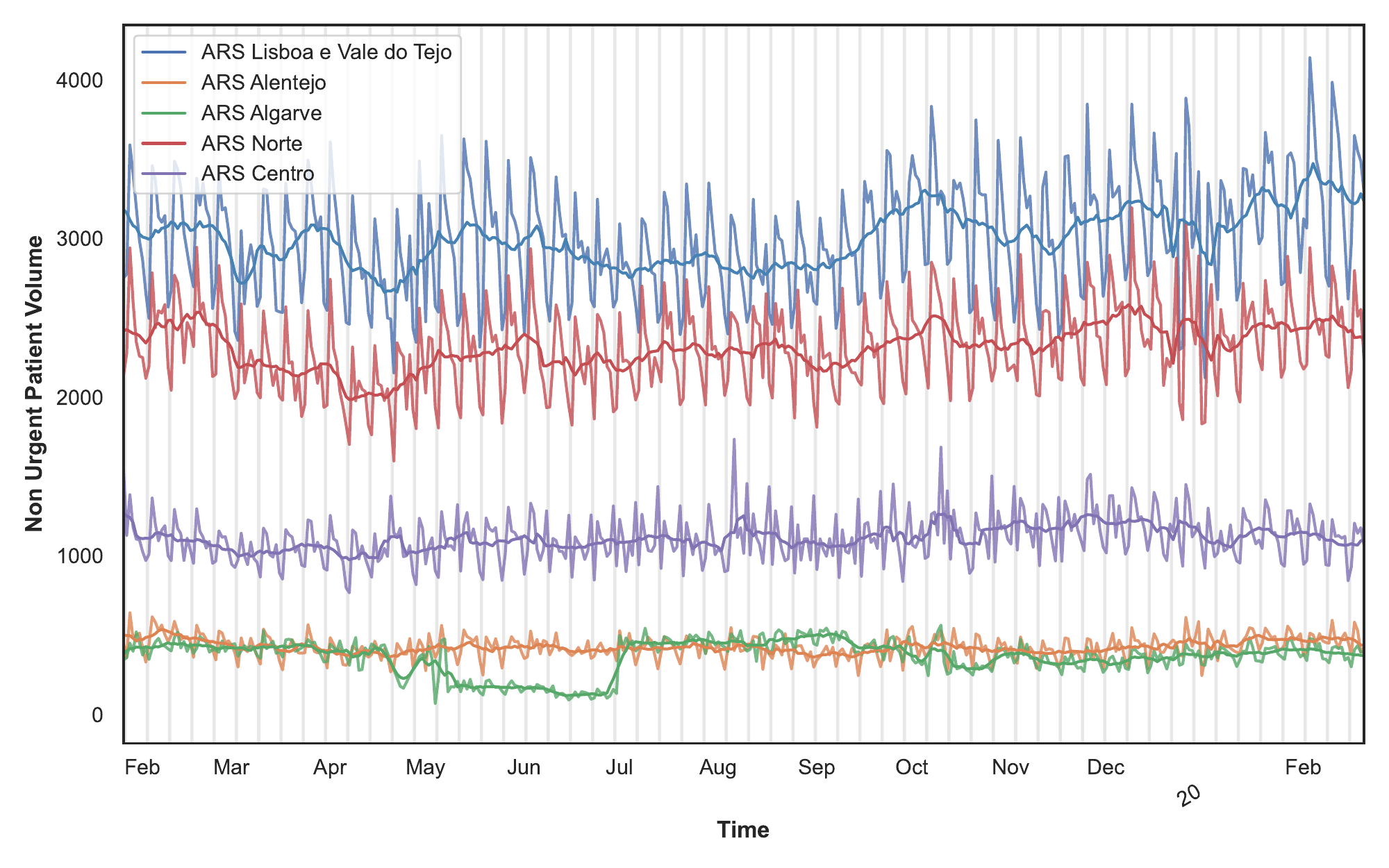}}
\end{figure}

In this section, we will present the dataset used in this work. The data was obtained from the public database “Transparência SNS”\footnote{\url{https://transparencia.sns.gov.pt/explore/dataset/atividade-sindrome-gripal-csh} \\ 
\url{https://transparencia.sns.gov.pt/explore/dataset/atendimentos-nos-csp-gripe}} and refers to daily data of care in primary health centres together with daily data of consultations and waiting times in hospitals’ emergency departments (ED) across Portugal, divided by Regional Health Area (RHA). The dataset covers the time period from November 1\textsuperscript{st}, 2016 to July 31\textsuperscript{th}, 2022; with 16 variables per observation that define, among other things, the Regional Health Area (RHA), \textit{Área Regional de Saúde (ARS)} in Portuguese, of the observation. In total, the dataset contains information regarding the daily volume of patients in emergency care, the number of scheduled and unscheduled consultations in primary health facilities, the daily number of patients arriving at the Emergency Department (ED) with respiratory issues, the waiting times between triage and the first medical evaluation and categorical variables pertaining to calendar information, such as weekend, day of week or national holidays.

In \figureref{fig:blue_white} we can observe the weekly variation in the number of non-urgent patients, as well as the volume shift according to RHA. It is visible that despite having different levels of affluence, the different RHA follow the same trend, with peaks of affluence occurring on Monday, and reduced volume on weekends and during the Summer months, usually associated with vacations. This aspect of the data served as motivation for the application of a non-linear model over multiple time series, unlike well-established models such as ARIMA.

\section{Study Setting}

Now that we have presented the data used in this paper, let us define, and expose, the reasoning behind the rules by which we will create and evaluate the model.

\begin{itemize}

    \item \textbf{Multivariate forecasting:} we want a model that leverages data and forecasts across different Regional Health Areas. In most research in the area, models are usually restricted to a single Hospital, and a more general model, capable of forecasting across different regions, might be able to uncover new interactions in the data and increase robustness.

    \item \textbf{Long-time forecast:} In order to add value at the hospital management level, the forecast of the number of patients should not be limited to the following day or week. In this paper, we have chosen a 4-week (28-day) forecast, considering that it allows breathing room for management and personnel decisions. To the best of our knowledge, few papers have worked on such an extended forecast horizon \cite{boyle2012,carvalho2018}, with only partial success.

    \item \textbf{Probability prediction:} besides obtaining an estimate of the most likely value in the future, a model that presents a probability density function on the prediction conveys much more information. Of special value is, for example, the definition of confidence intervals, which can transmit to those who use the model an idea of the confidence, or precision, of the model in its estimation.% Almost all classical linear methods, such as ARIMA or Exponential Smoothing, are able to deliver confidence intervals over the predictions. However, the same is not true for common Neural Networks architectures.

    \item \textbf{Explanatory variables:} Importantly, we want to evaluate the predictive capacity of different variables, determining up to which passed time-step the model finds predictive value or which covariates, categorical or numerical, have a significant impact on the prediction.

\end{itemize}

\section{Models}
\label{sec:mod}

The first and simpler method used for forecasting is the replication of the last $k$ time-steps. This technique, which is used as Baseline in this paper, is also referred to as the naïve algorithm. By evaluating this model on the validation, the optimal value for $k$ was estimated to be $7$, thus representing the weekly periodicity in the data.

For comparison, other models commonly used in this area were also applied, namely AutoRegressive Integrated Moving Average (ARIMA) with a seasonal component \citep{Jones2008,Schweigler2009,Eyles2022}, and its multivariate variant Vector AutoRegressive Integrated Moving Average (VARIMA) \citep{Kadri2014}. 

Also used was the exponential Smoothing algorithm, a simple method that has also shown good results in the literature \citep{Champion2007}. Finally, to gauge the performance of common machine learning models, the XGBoost model was used. Out of these models, the XGBoost \citep{NIPS2017_6449f44a} (a Decision Tree Boosting algorithm), is the only model capable of using past and future covariates, with the disadvantage of not being specifically tailored for time-series data. All these baselines models have statistical variants, or extensions, that can estimate a prediction interval, or otherwise a probability distribution that can in turn be used to define prediction intervals.

Initially introduced by \citet{Lim2021}, the Temporal Fusion Transformer (TFT) model instantiated a novel architecture, combining a few mechanisms previously only used separately, in a single model. The key features of the TFT are: 

\begin{itemize}
    \item Variable Selection Network: three independent Selection Networks, one for each variable set (static, past and future), to select only relevant variables at each time-step. This module removes noisy variables that do not add predictive value, while giving some level of insight into the variables that are more significant to the prediction;
    \item A Gating Mechanism to skip any other element of the architecture. For specific cases where exogenous variables are not useful or there is no need for non-linear processing (e.g. in very simple forecasts), the Gating Mechanism, also referred to as Gated Residual Network \citep{He_2016_CVPR}, allows the model to only use non-linear processing when needed;
    \item Static variables encoding to combine static information with time-series data;
    \item Temporal Dependency Processing to capture short-term dependency, with an LSTM encoder-decoder \citep{Fan2019}, and long-term dependency using a Multi-Head Attention mechanism \citep{Vaswani2017}. By an additive aggregation of the different heads, this mechanism gains explainability, as the weights in the aggregated Multi-head represent time-step importance;
    \item Confidence Intervals: the output of the models are quantiles, that define prediction intervals, at each forecast time-step.
\end{itemize}

The hyperparameters of the model, alongside a more in-depth explanation of its features and a diagram of its architecture, are detailed in Annex \ref{chap:models}.

\section{Results}

In this section, we present the results of the TFT and five other models for a 4 week forecast window over a 6 months period of time, from January, 24 to July, 31 2022. \tableref{tab1} illustrates how the TFT outperforms other common models in the literature for long time prediction, with a Root Mean Squared Error (RMSE) of 178.19, or approximately 178 people per day. This metric, however, might be deceptive, as it is scale dependent, meaning that RHAs with a larger daily volume will necessarily yield a higher RMSE, and skew the results. The Mean Absolute Percentage Error (MAPE) on the other hand, is scale invariant, and it better depicts the overall predictive power of the models, with the TFT obtaining a 9.87\% percentage error. To better compare the models, we utilised an empirical CDF for each model, as seen in  \figureref{fig:ecdf}. In this Figure, depicting Absolute Error, the TFT shows overall slightly better performance. We also acknowledge that the VARIMA algorithm obtains somewhat good results and that the Gradient Boosting Tree obtains a better score with its 95\% confidence interval, however this is in part due to the fact that no observation falls outside of the confidence, which might indicate that the interval is to wide. By analyzing the results, the authors also found that the statistical methods, such as ARIMA or Exponential Smoothing, had degraded results in the last 2 weeks of the forecast window, unlike the Gradient Boosted Tree method and the TFT. In Table \ref{tab2} this effect can be observed by evaluating the MAPE error per week.

% nao esquecer de por tabela
This illustrates the superior capability of machine learning to perform long term prediction, as the complexity of the model helps to identify long term patterns, and identify break points where the trend shifts.

%\begin{table}[t!]
%\center
%\caption{Prediction accuracy for various models in the period %24/01/2022 -- 20/02/2022. To evaluate the models, four %metrics are used: Mean Absolute Error (MAE), Root Mean %Squared Error (RMSE),Mean Absolute Percentage Error (MAPE), %and Mean Squared Error (MSE). Bold indicates the best %result.}
%\label{tab1}
%\hfill
%\center
%\begin{tabular}{c c c c c c}
%\toprule
%Models & MAE & RMSE & sMAPE & MAPE & MSE \\
%\midrule
%Baseline & 95.1643 &116.5850  &8.3236 &7.3483 & 20245.0643 \\
%Exp. Smoothing & 112.5885 &135.6158 & 8.3835 & 7.3135 & %29888.6468\\
%ARIMA &104.9886 &129.6084 &8.8641 &7.8484 & 22949.7471 \\
%VARIMA & 94.6441  &120.6674 & 8.6732 & 7.9250 & 18407.9554 \\
%XGBoost  &92.0307 & 112.3295 & 9.0277  &  7.7027 & %16178.5531\\
%TFT & \textbf{66.7551} &\textbf{84.4102} &\textbf{7.0817} & %\textbf{5.9084}& \textbf{8379.7340}  \\
%\bottomrule
%\end{tabular}
%\end{table}

\begin{table}[hbtp]
\setlength{\tabcolsep}{1.3pt}
\floatconts
{tab1}
{\caption{Forecast error in the  24/01/2022 -- 31/07/2022 period. To evaluate the models, five metrics are used: Mean Absolute Error (MAE), Root Mean Squared Error (RMSE), Mean Absolute Percentage Error (MAPE) ,Mean Squared Error (MSE) and Mean Interval Score (MIS). Bold indicates the best result; TFT is consistently more accurate than the baselines.}}
{\begin{tabular}{c c c c c c}
\toprule
 Models &          MAE &  RMSE & MAPE & MSE &  MIS \\
\midrule
\small{Baseline} & 160.85 &192.01   &11.14 & 56615 & \-- \\
\small{Exp. S.} & 189.32 & 222.71  & 12.40 & 83522 & 1114\\
\small{ARIMA} &160.48 &189.79  &10.60& 61234 & 2848\\
\small{VARIMA} & 159.01  &187.43  & 10.32 & 56776 & \--\\
\small{XGBoost}  &192.48 & 233.34   &   12.17 &  103569 & \textbf{730}\\
\small{TFT} & \textbf{151.60} &\textbf{178.19} & \textbf{9.87}& \textbf{55559} & 768 \\
\bottomrule
\end{tabular}}
\end{table}

\begin{figure}[htbp]
\floatconts
  {fig:ecdf}
  {\caption{Empirical Cumulative Distribution function for the absolute difference between the true value and the predicted value, for all RHA.}}
  {\includegraphics[width=\linewidth]{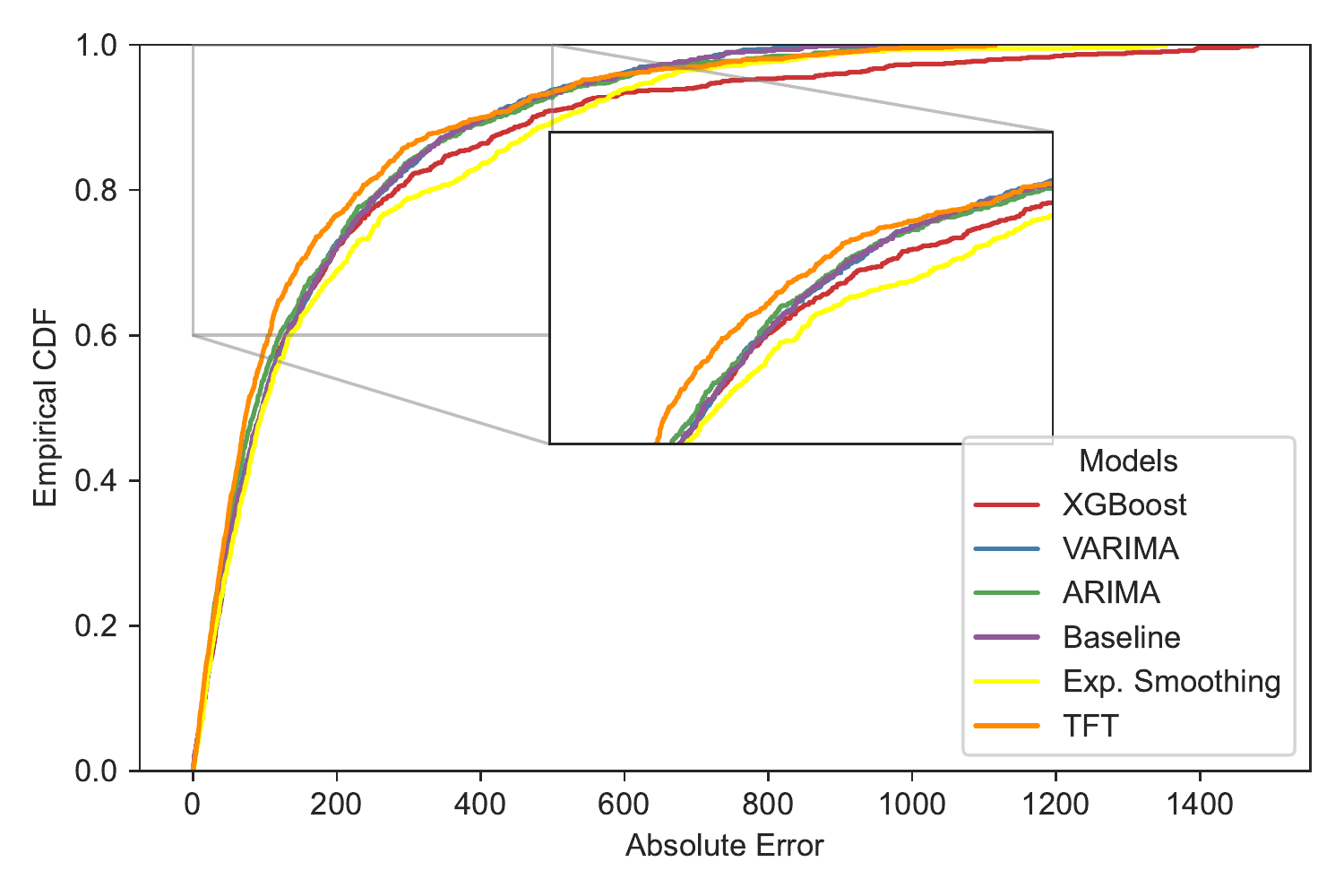}}
\end{figure}

But the strength of the Temporal Fusion Transformer used goes beyond the precision of the model. The ability to give some insight into how the forecast was obtained is just as important. Two key findings can be summarized from the results: first, we can observe the attention given to each time-step and determine that the most recent time-steps are given in average a higher weight, which is intuitively expected and shows that old information has less value to the model. Secondly, the two most relevant covariates are the percentage of patients in the emergency room with respiratory problems and holidays. It is possible that an increase in the percentage of patients with respiratory problems might be indicative of a spike in influenza/COVID-19 transmission, thus indicating future positive trends in the number of non-urgent cases. During holidays the number of patients using the emergency department drops significantly, so we would expect the model to give particular importance to this one feature. A more detailed analysis of the covariates and the accompanying Figures is presented in Annex   \ref{apd:second}.

\begin{table}[hbtp]
\setlength{\tabcolsep}{1.05pt}
\floatconts
{tab2}
{\caption{MAPE error by week; Exponential Smoothing showed the best results in the first week, followed by quick degradation for longer forecast periods.}}
{\begin{tabular}{c c c c c c}
\toprule
 Models &          week 1 &	week 2 &	week 3	& week 4	&Total \\
\midrule
\small{Baseline} & 7.16 &	10.01 &	12.84 &	14.55	 & 11.14 \\
\small{Exp. S.} & \textbf{5.72} &	10.30 &	14.95 &	18.65	& 12.40\\
\small{ARIMA} &8.57 &	\textbf{9.40} &	11.72	&12.70	& 10.60\\
\small{VARIMA} & 8.09	&9.47&	11.22&	12.50 &	10.32\\
\small{XGBoost}  &10.64 &	11.94	&12.44&	13.66	&12.17\\
\small{TFT} & 8.81 &	9.91&	\textbf{9.88}	& \textbf{10.82}&	\textbf{9.87} \\
\bottomrule
\end{tabular}}
\end{table}

\section{Conclusion}

This paper presented a novel application of the Temporal Fusion Transformer (TFT) model to predict non-urgent patient volume in Portuguese public hospitals by Health Regional Areas (HRA). The results were encouraging, surpassing other models commonly found in the literature \citep{Jones2008,Kadri2014}. The forecasting of an entire month is seldom done in the literature \citep{carvalho2018,abo2015}, and it could be further enhanced by a thorough analysis of model deterioration over time.
The introduction of a multivariate model with good results across groups is a positive prospect, since one limitation of univariate time-series is the natural low-data regimen, while multivariate models can merge information from multiple sources, thus increasing the total amount of data fed to the models. In the future, this model can increase in granularity, forecasting at the hospital level instead of aggregated values by HRAs. Although a greater challenge, due to the increased noise and randomness that comes from the decrease in the study population, we expect that the combination of a large number of time-series could improve the robustness and global quality of the model, specifically if we add more relevant static variables. T For this paper, only HRA and time-series statistics were used as static covariates, but as noted in \cite{Farmer307}, across different regions there is distinct demand for emergency care, thus impacting the scale and variance of the time-series. In future work, we plan to introduce other factors that might contribute to encode region-specific information as static covariates, such as demographics, modes of transport available, socio-economic characterisation of the patient population and number and capacity of private healthcare providers in the region. All these elements might help to represent each class, and ultimately be used for a generalisation of the model to unseen hospitals, where these variables might help to represent how similar a new unseen hospital/RHA is to hospitals/RHAs in the training data.

\bibliography{pmlr-sample}

\appendix

\newpage
\section{Literature Review, Data and Models}\label{apd:first}

\subsection{Literature Review}

As mentioned before in the main section of the paper, much work as been done using ARIMA and other statistical models to forecast ED overcrowding, with more recent work using Deep Learning techniques suited to time series such as RNNs or LSTM \citep{Kadri2020,Harrou2020}, and many works pointing to the impact of calendar variables \citep{boyle2012,Jones2008,Wargon2009} in improving forecasts. Weather data, such as temperature and rain, has also shown predictive power for ED arrivals with respiratory problems \citep{Navares2018}, but in others studies that analysed the whole spectrum of ED visitors, it is either not a significant variable, or it could be replaced by calendar variables, e.g. month of the year \citep{hertzum2017,Wargon2009}. 
Covariate interpretability is one of the frequent drawbacks of Neural Networks, alongside the failure to recognize long-term dependencies in time-series. One specific device that addresses both problems is the Attention Mechanism \citep{Vaswani2017}: simply put, it evaluates long-term dependencies and also represents how each time-step impacts the model’s prediction. Attention has been used as part of a specific Neural Network family of architectures called Transformers, that has shown impressive results in the Natural Language Processing field \citep{devlin-etal-2019-bert,wolf-etal-2020-transformers}. In the literature, we found only one example that used a Temporal Fusion Transformer model to predict Emergency Department (ED) volume in one hospital for one day ahead \citep{Pulkinnen2020}.
While not being the only work that performed only daily predictions \citep{Rocha2021,Tuominen}, we find that a longer forecasting window produces increased value for hospital management and poses a different challenge from a machine learning perspective, as seasonal fluctuation needs to be fully represented, and common forecasting models tend to decrease in predictive quality as the forecast period becomes wider.

\subsection{Data Analysis}
\label{chap:dataa}
Due to the time period and the type of data, it is important to consider the Covid Lockdown period. Portugal's response to the Covid-19 Pandemic resulted in a series of containment measures that reduced travel and in person work: in this period (10/03/2020 -- 1/08/2021) the percentage of non urgent visits in the RHA of \textit{Lisboa e Vale do Tejo} (Lisbon and Tagus Valley) dropped from the normal value of 48\% of the total to 40\%, with more dramatic drops for example in the \textit{Algarve} RHA from 45\% to 30\% at the beginning of the pandemic. This dramatic period influenced the way people used emergency services, and it can demonstrate how external factors influence people going to the emergency room. This shift, associated with the general decline in the number of people in urgent care, represents a distribution change in the time series, therefore making it exceedingly difficult to predict the COVID period using only pre-COVID information. In the same way, we can conclude that this COVID period does not have useful information about the post-COVID future, and, in fact, we have experimentally verified that the quality of the models decreased with the introduction of the COVID period, thus leading to the decision to exclude this temporal section from the training set.

It is, in a certain way, clear that the prediction of the influx in emergency rooms can be useful for a more efficient management of hospital services, but there is visible value added at user level, in the sense that they will get better and faster care \citep{pines2008}. To sustain this claim, we can observe the impact that the number of non-urgent patients has on the waiting time before being treated in the Emergency Department. In \figureref{fig:time_blue}, for the RHAs with the highest daily affluence, we observe a positive, moderate to strong correlation between the number of non-urgent patients and the waiting time. This is an indicator, not entirely unexpected, that ED overcrowding of non urgent patients can lead to a substantial increase in the average waiting time for all patients, urgent or non urgent.

\begin{figure}[htbp]
 % Caption and label go in the first argument and the figure contents
 % go in the second argument
\floatconts
  {fig:time_blue}
  {\caption{Correlation between waiting times and non urgent patient volume. A strong to moderate correlation exist between these two variables, therefore implying that overcrowding increases waiting times.}}
  {\includegraphics[width=0.99\linewidth]{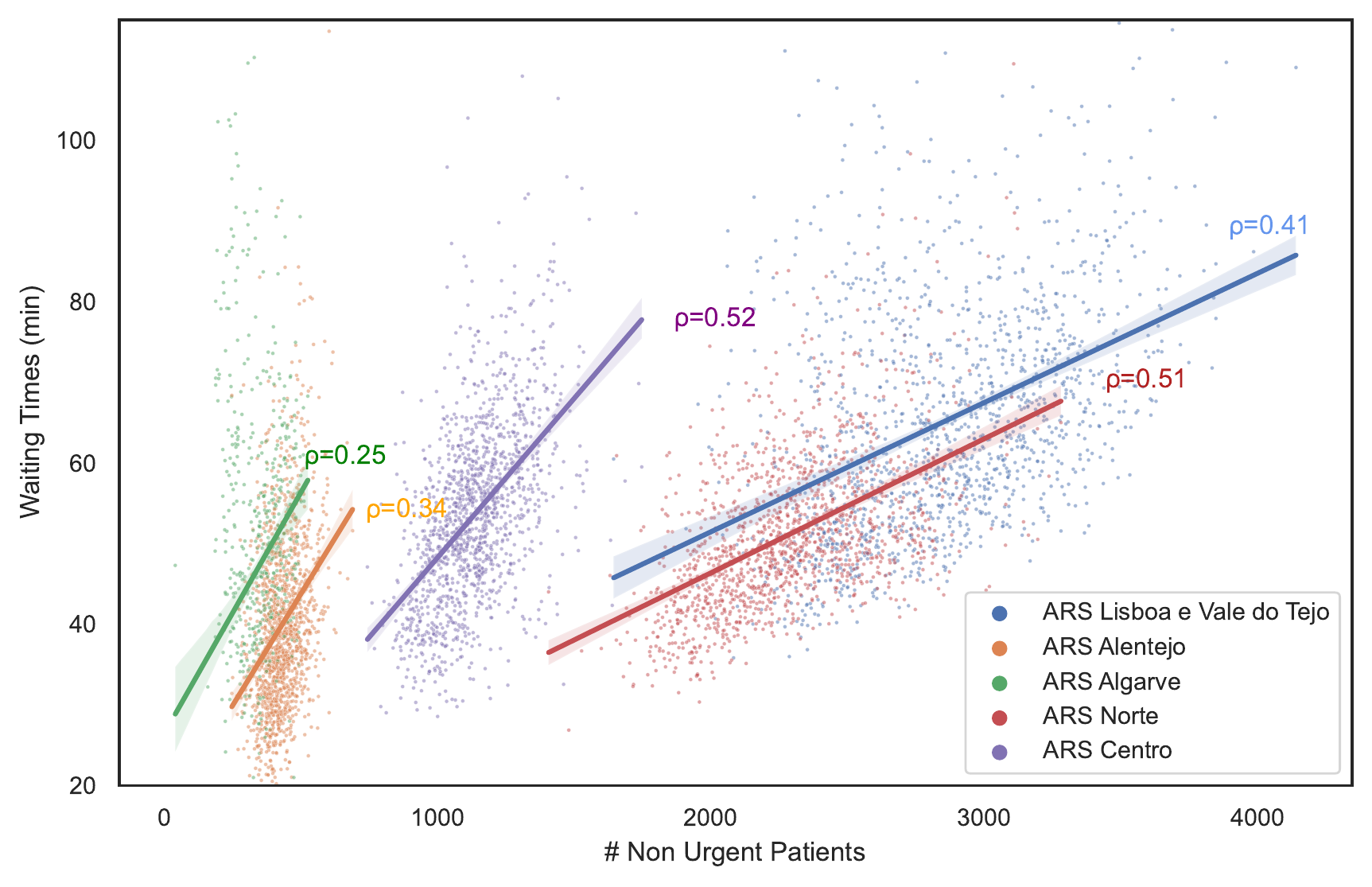}}
\end{figure}

%, which lead to improved results. 

\subsection{Metrics}
\label{sec:meth}
%In most research in the area to this day, two issues occur commonly: a regional specificity of data, where data is usually from one or more Hospitals in a given area, and short forecasting, where data is only forecast to the next day, or next week. To address these limitations, we will now define the goals by which we choose the forecasting methods:

In this paper, we use five metrics to evaluate the models. The Mean Absolute Error (MAE):
\begin{equation}
    \text{MAE} = \frac{1}{5} \sum_{k=1}^5 \frac{1}{n}\sum_{i=1}^n |y_i^k-\hat{y}_i^k|
\end{equation} 
the Root Mean Squared Error (RMSE),
\begin{equation}
    \text{RMSE} = \frac{1}{5} \sum_{k=1}^5 \sqrt{\frac{1}{n}\sum_{i=1}^n (y_i^k-\hat{y}_i^k)^2}
\end{equation} 
the Mean Absolute Percentage Error (MAPE),
\begin{equation}
    \text{MAPE} = \frac{1}{5} \sum_{k=1}^5 \frac{1}{n}\sum_{i=1}^n \frac{|y_i^k-\hat{y}_i^k|}{|y_i^k|}
\end{equation} 
the Mean Squared Error (MSE),
\begin{equation}
    \text{MSE} = \frac{1}{5} \sum_{k=1}^5 \frac{1}{n}\sum_{i=1}^n (y_i^k-\hat{y}_i^k)^2.
\end{equation} 
and finally the Mean Interval Score (MIS) \citep{Gneiting2007} to evaluate the confidence interval with $\alpha=0.95$,
\begin{align}
    \text{MIS}_\alpha =& \frac{1}{5} \sum_{k=1}^5 \frac{1}{n}\sum_{i=1} \big( (\hat{U}_i^k - \hat{L}_i^k)  \nonumber\\ 
    & + \frac{2}{\alpha}(\hat{L}_i^k - \hat{y}_i^k) \mathbbm{1}(y_i^k < \hat{L}_i^k)   \nonumber \\
    & + \frac{2}{\alpha}(\hat{y}_i^k - \hat{U}_i^k) \mathbbm{1}(y_i^k > \hat{U}_i^k)\big),
\end{align}
where $\hat{L}_i^k$ and $\hat{U}_i^k$ are, respectively, the predicted lower and upper bound and $\mathbbm{1}$ is the characteristic function. This score rewards a small confidence interval, while applying a penalty whenever the predicted result falls outside of the interval.
Since we are evaluating the predictions over several groups (RHAs), the total error will be the average across RHAs. The most common metric across the literature for ED forecasting is the Mean Absolute Percentage Error (MAPE) \citep{carvalho2018,EKSTROM2015436}, however, when the true value is close to zero, this metric becomes unreliable. It also places a heavier penalty on negative errors (when the predicted value is higher than the true value) \citep{MAKRIDAKIS1993527}. To overcome that, outliers values very close to zero are removed for this particular metric. 
To correctly evaluate the out-of-sample predictive capacity of the model, the dataset is divided into three subsets: train, validation, and test. The training and validation sets combined represent roughly 4 years, while the test set is roughly 6 months. The validation set is used to optimise hyperparameters and to identify overfitting during training, while the test set is unseen until the end and is only used to produce the final results.

\subsection{Temporal Fusion Transformer}
\label{chap:models}
\begin{figure}[htbp]
 % Caption and label go in the first argument and the figure contents
 % go in the second argument
\floatconts
  {fig:tft}
  {\caption{TFT architecture. The inputs are static metadata, time-varying past inputs (including past target values) and known future information. The Variable selection unit selects the most relevant features, while the Gated Residual Network allows to skip over unused sections of the architecture. The interpretable multi-head attention is used to evaluate the most relevant time-steps. Image adapted from \citet{Lim2021}.}}
  {\includegraphics[width=0.9\linewidth]{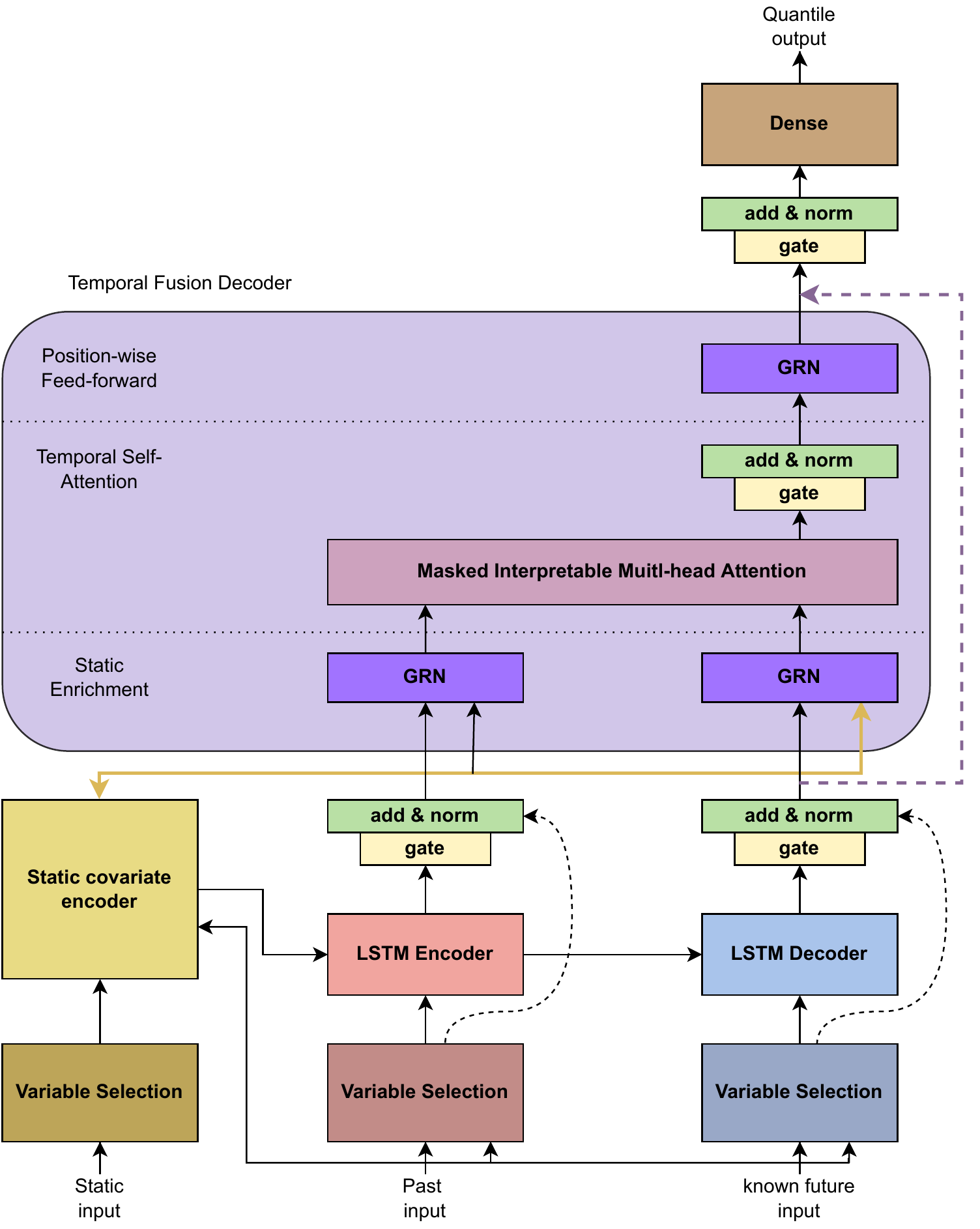}}
\end{figure}
%The first and simpler method used for forecasting is the replication of the last $k$ time-steps. This technique, which is used as Baseline in this paper, is also referred to as the naïve algorithm. By evaluating this model on the validation, the optimal value for $k$ was estimated to be $7$, thus representing the weekly periodicity in the data.

%For comparison, other models commonly used in this area were also applied, namely AutoRegressive Integrated Moving Average (ARIMA) with a seasonal component \cite{Jones2008,Abdel_Aal_1998,Schweigler2009,Kadri2014,Eyles2022}, and its multivariate variant Vector AutoRegressive Integrated Moving Average (VARIMA) \cite{Kadri2014}. 

%Also used was the exponential Smoothing algorithm, a simple method that has also shown good results in the literature \cite{Champion2007}. Finally, to gauge the performance of common machine learning models, the XGBoost model was used. Out of these models, the XGBoost \cite{NIPS2017_6449f44a} (a Decision Tree Boosting algorithm), is the only model capable of using past and future covariates, with the disadvantage of not being specifically tailored for time-series data. All these baselines models have statistical variants, or extensions, that can estimate a prediction interval, or otherwise a probability distribution that can in turn be used to define prediction intervals.

The novel model used in this paper is the Temporal Fusion Transformer. In Section \ref{sec:mod} of the main paper the key features of the model are defined, starting with the Variable Selection Network. To understand this mechanism, we first need to define the model input, and separate the input variables into static, target and time-dependent. Static covariates, such as time-series variance or mean, are specific to each group, i.e. RHA, and are defined as $s_i$ with $i={0,...,4}$. $y_{i,t}$ is the target for group $i$ at time-step $t$ and $x_{i,t} = [p_{i,t}^T,f_{i,t}^T]^T$ the time-dependent covariates, with $p$ representing past covariates, meaning covariates that are only known until the present, and $f$ future covariates, that can be assumed to be known in the past and the future, in our case, holidays and weekends. The Variable Selection Network is independent for each type of input, and it produces feature weights that will be used to interpret the results, with higher weights ascribed to the most important covariates.
The Masked Interpretable Multi-head Attention is used to learn long-term relationships between different time-steps and to determine which time-steps are more relevant to the model prediction. Multi-head means that multiple attention mechanisms are applied over the sequence, and it was presented by \cite{Vaswani2017} to improve the learning capacity of the standard attention mechanism, however, the use of multiple heads over different sections of the input sequence hinders interpretability. A solution presented by the authors of the TFT model \citep{Lim2021} is to use multiple heads over the same sequence, and then ensembled them into a single interpretable head, thus combining the learning capacity of multi-head attention with the explainability of a single attention mechanism.

The prediction function is defined as \citep{Lim2021}:

\begin{equation}
   \hat{y}_i(q,t,\tau) = f_q(\tau,y_{i,t-k:t},x_{i,t-k:t},s_i)
   \label{eq:pf}
\end{equation}

where  $\hat{y}_i (q,t,\tau)$ is the predicted $q$th quantile for the $\tau \in \{1,..,\tau_{max}\}$ value in group $i$, at time $t$. For the specific case of this work, $\tau_{max} = 28$, as we want to forecast simultaneously $28$ days ahead. By predicting quantiles, we obtain a quasi-distribution of the expected value, and gain the capacity to define confidence intervals.
\begin{table}[t!]
    \centering
    \caption{Hyperparameters for TFT model after tuning.}
    \begin{tabular}{lr}
    \toprule
    hyperparameter & value \\
    \midrule
        encoder length & 42   \\
          batch size & 40 \\
           prediction length & 28  \\
           gradient clipping & 0.022730 \\
           learning rate & 0.0011149 \\
           hidden size & 33 \\
           number of attention heads & 8 \\
           dropout & 0.19230 \\
           hidden continuous size & 19 \\
\bottomrule
    \end{tabular}
    \label{tab:hyp}
\end{table}
%Initially introduced by \cite{Lim2021}, this model instantiated a novel architecture, combining a few mechanisms previously only used separately, in a single model. The key features of the TFT are: 

%\begin{itemize}
 %   \item Variable Selection Network: three independent Selection Networks, one for each variable set, to select only relevant variables at each time-step. This module removes noisy variables that don’t add predictive value, while giving some level of insight into the variables that are more significant to the prediction;
 %   \item A Gating Mechanism to skip any other element of the architecture. For specific cases where exogenous variables are not useful or there is no need for non-linear processing (e.g. in very simple forecasts) the Gating Mechanism, also refereed to as Gated Residual Network  \cite{He_2016_CVPR}, allows the model to only use non-linear processing when needed;
%    \item Static Variables encoding to combine static information with time-series data;
 %   \item Temporal Dependency Processing to capture short-term dependency, with an LSTM encoder-decoder \cite{Hochreiter1997,Fan2019}, and long-term dependency using a Multi-Head Attention mechanism \cite{Vaswani2017}. By an additive aggregation of the different heads, this mechanism gains explainability, as the weights in the aggregated Multi-head represent time-step importance;

%    \item Confidence Intervals: the output of the models are quantiles, that define prediction intervals, at each forecast time-step.
%\end{itemize}

To obtain the quantile predictions, a specific loss, the Quantile Loss, is defined as \citep{koenker2001quantile}:

\begin{equation}
    QL(y,\hat{y},q) = \max\{q(y-\hat{y}),(q-1)(y-\hat{y}) \}
\end{equation}
for each quantile $q$. The final Loss is the average QL across quantiles and for the entire prediction horizon [0,$\tau_{max}$]. In this work, the quantiles used were [0.2,0.1,0.25,0.5, 0.75,0.9,0.98]. When $q=0.5$ the Loss is equal to MAE divided by 2, and $q=0.5$ (the median) is the value used for the point-wise prediction of the model. The complete hyperparamenters of the model are defined in \tableref{tab:hyp} and the
overall architecture of the TFT can be seen in \figureref{fig:tft}.

\section{Results}
\label{apd:second}

\begin{figure}[htbp]
 % Caption and label go in the first argument and the figure contents
 % go in the second argument
\floatconts
  {fig:pred}
  {\caption{Test sample forecast. Over the input vector, we can see the grey line representing attention. In orange is the median predictive value (q=0.5), with different quantiles shown as shaded area.}}
  {\includegraphics[width=0.9\linewidth]{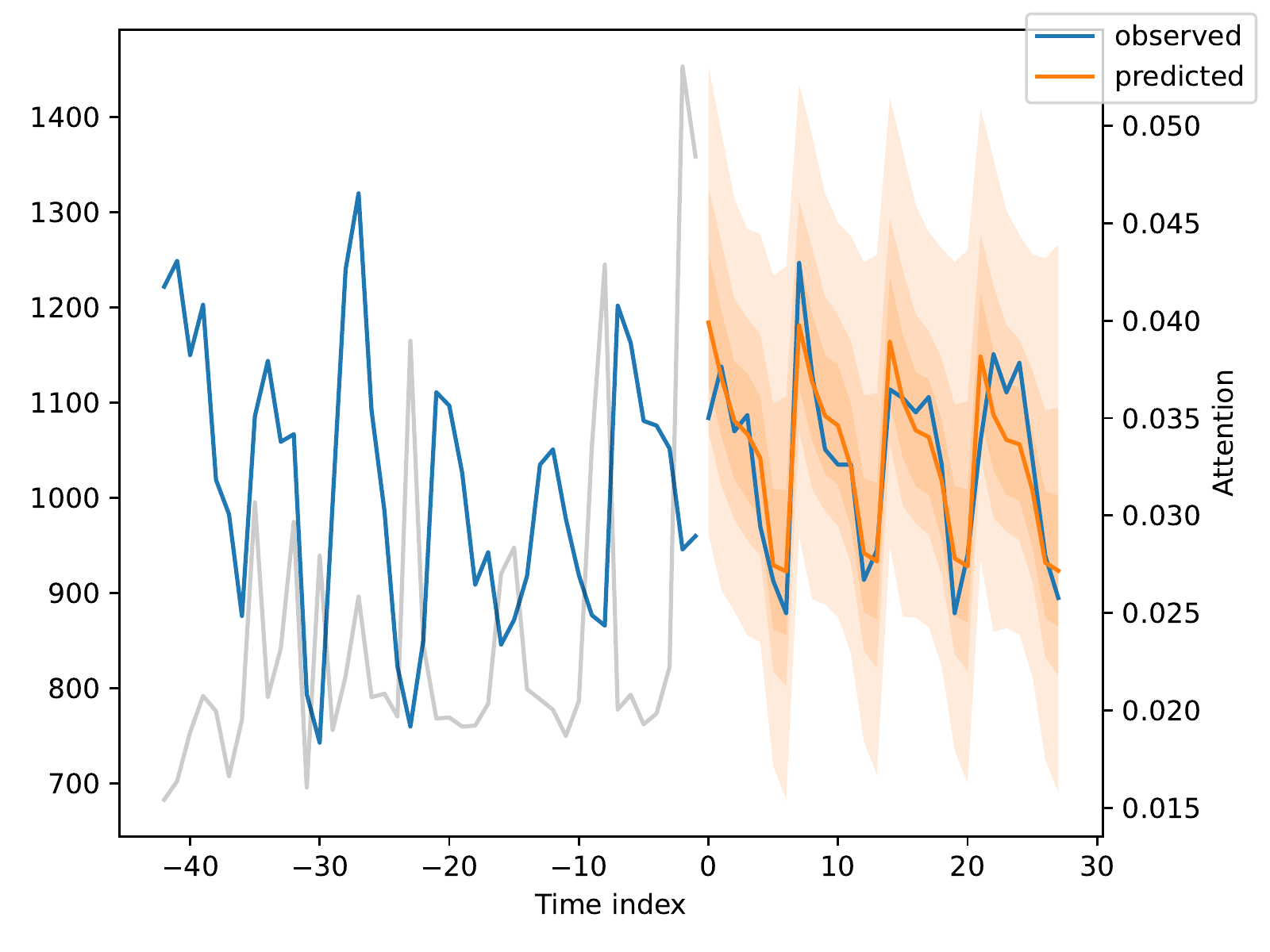}}
\end{figure}

The two elements of explainability that are given by the model are time-step attention and variable importance.
In Figure \ref{fig:pred} we have a sample forecast from the test set. In grey are the attention weights given to each time-step. These values change according to the input, so if we average the attention for all predictions we can get a sense of the most relevant time-steps. In Figure \ref{fig:att}, it can be distinguished how the model values the most recent time-steps with a higher weight, which is intuitively expected and shows that old information has less value to the model. This validates a common assumption in linear models, that ascribe more weight to more recent observations, as is the case of the Exponential Smoothing model. In addition, we observe another more intriguing feature, already observable in the sample forecast, which are spikes in attention during the weekend; these spikes may occur because particular attention is given to one or two previous weekends to define patient volume in future weekends.

\begin{figure}[htbp]
 % Caption and label go in the first argument and the figure contents
 % go in the second argument
\floatconts
  {fig:att}
  {\caption{Average attention attributed over the input vector. More recent time-steps have higher values.}}
  {\includegraphics[width=0.96\linewidth]{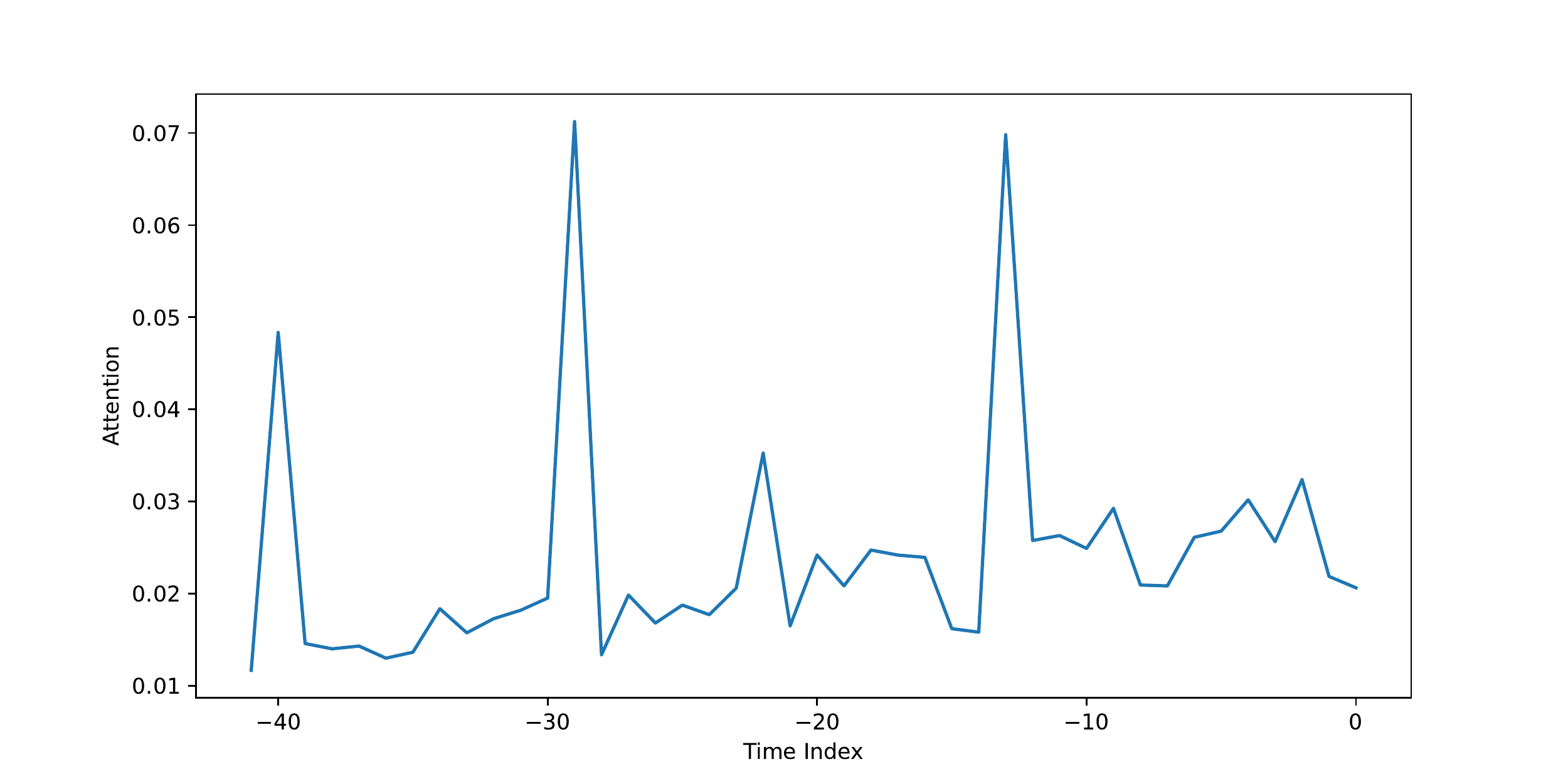}}
\end{figure}

The model categorises covariates into three types: static, past, and future. And, as a consequence of using the Variable Selection Network, features can be listed by importance. 
In \figureref{fig:previsoes,fig:previsoes2} we have, respectively, the past and future covariates importance. The most relevant past feature is the percentage of patients in the emergency room with respiratory problems. For this period, excess affluence in hospital emergency rooms could be attributed to peaks in influenza/COVID-19 transmission, it therefore makes sense that this variable can be a predictive indicator of future positive trends in the number of non-urgent cases.  The second most important variable is patient waiting time, which is in line with the positive relationship presented in \sectionref{chap:dataa} between the increase in waiting time and the increase in non-urgent patients. However, we should not focus our attention solely on the variables relevant to the model. There is interest in observing the variables that did not add value to the model; here we can observe that the information regarding healthcare centres (\textit{n\_cons\_total, prog}) did not add value to the model, meaning that there is no clear interaction between patient volume in healthcare centres, mostly used for primary healthcare and minor health issues, and non-urgent patients in Emergency Departments.
%this case, we can observe that the total number of patients in the emergency room has a marginal importance for the model. Although the interpretation of this result is more difficult, we can propose a possible interpretation: The total number of patients in the emergency room is not relevant to predict the future, because urgent patients (yellow grade to higher) go to the emergency room independently of non urgent patients.

The most important future covariate feature is the categorical variable indicating public holidays in Portugal. The model has attributed such an importance to holidays because they have a severe impact on patient volume, not only on the day, but also on the next day, when close to the weekend. Furthermore, the other future covariates have a non-negligible importance both as past and future covariates, thus supporting the claim found in the literature that calendar variables have a significant impact on the prediction. Although these elements of explainability certainly add value to the model, there are still some limitations when interpreting these results. In the case of the variable importance, it is not possible to ascertain if the impact of each variable is positive or negative. Furthermore, because these three networks are separate, it is not possible to compare past with future variables, nor the total impact of a given input type (static, past or future).

\begin{figure}[htbp]
 % Caption and label go in the first argument and the figure contents
 % go in the second argument
\floatconts
  {fig:previsoes}
  {\caption{Past Covariates Importance. The most relevant past feature to the model is the percentage of patients in ED  with indication of respiratory problems.}}
  {\includegraphics[width=1.1\linewidth]{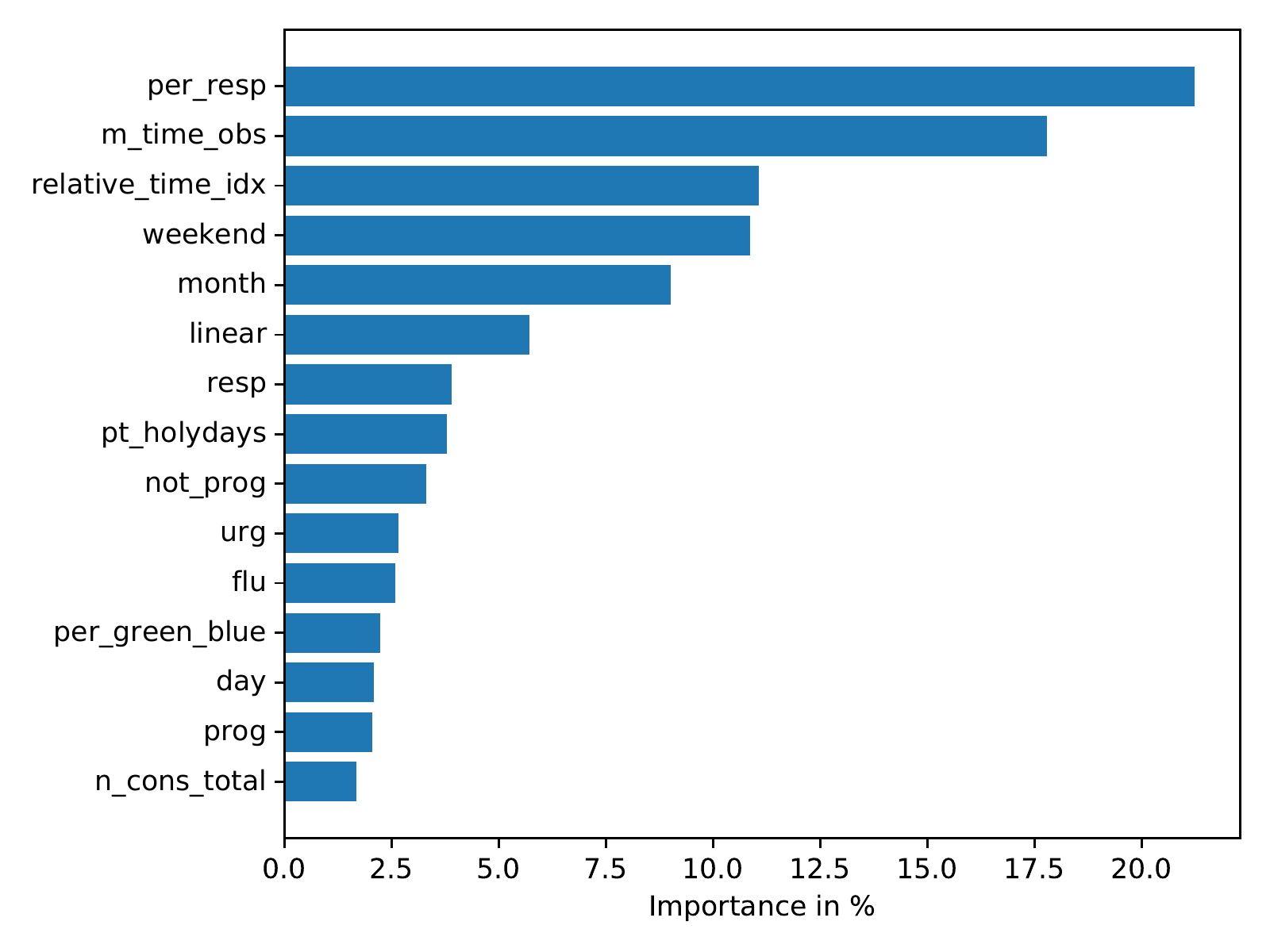}}
\end{figure}

\begin{figure}[htbp]
 % Caption and label go in the first argument and the figure contents
 % go in the second argument
\floatconts
  {fig:previsoes2}
  {\caption{Future Covariates Importance.  For future
covariates, variables that are known in the future, the most relevant one is a feature
that indicates public holidays in Portugal.}}
  {\includegraphics[width=1.1\linewidth]{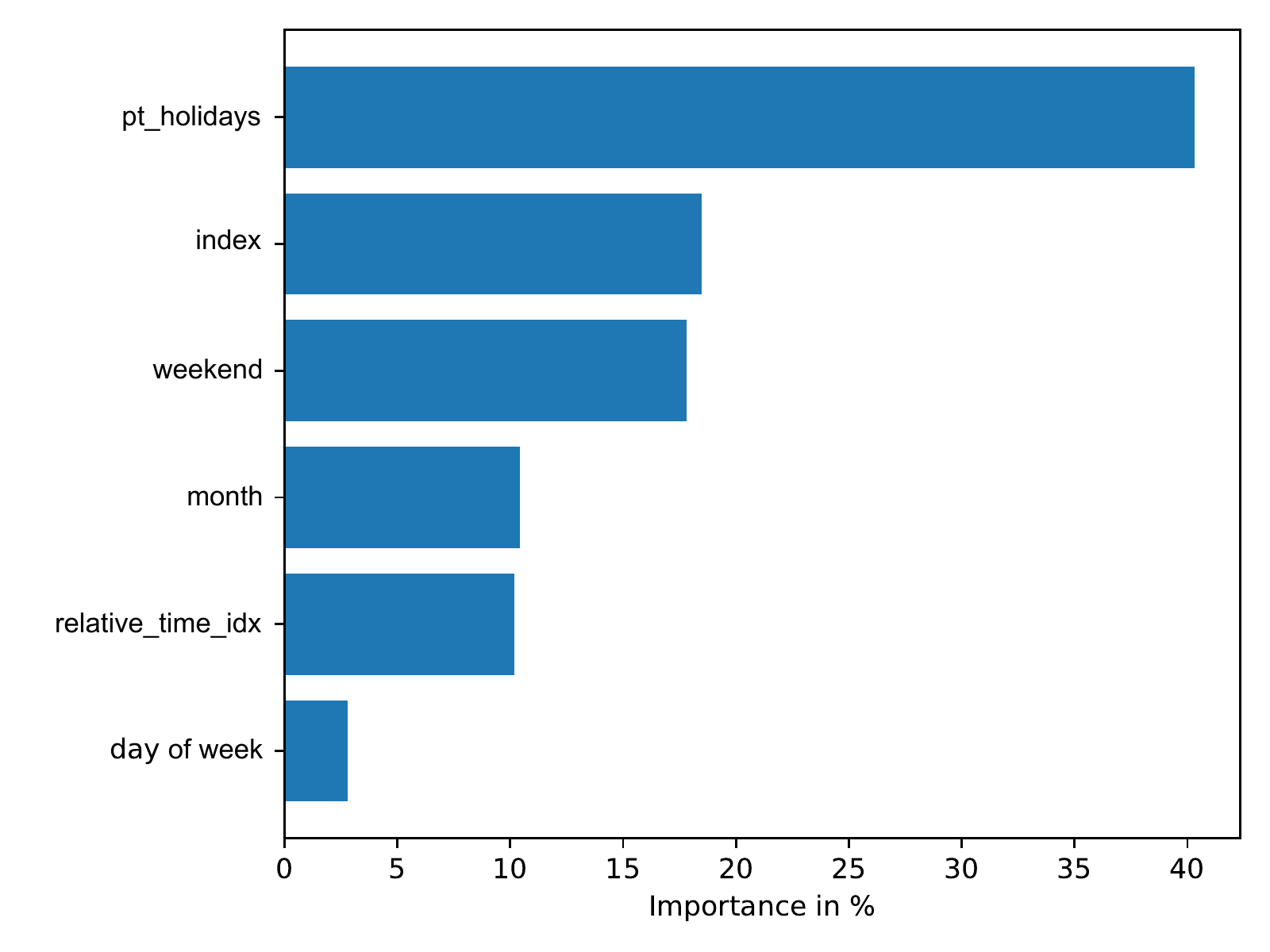}}
\end{figure}

\end{document}